\documentclass[aps,pre,twocolumn,groupedaddress,showpacs]{revtex4}
\usepackage{graphicx}
\newcommand{\beq}{\begin{equation}}
\newcommand{\eeq}{\end{equation}}
\newcommand{\beqn}{\begin{eqnarray}}
\newcommand{\eeqn}{\end{eqnarray}}
\newcommand{\pp}{\partial}

\begin{document}


\title{Nonequilibrium dynamics of polymer translocation: a mean-field 
model}
\author{Chiu Fan \surname{Lee}}\email{C.Lee1@physics.ox.ac.uk}
\affiliation{Physics Department, Clarendon Laboratory, 
Oxford University, Parks Road, Oxford OX1 3PU, UK}

\date{\today}

\begin{abstract}
We analyse the dynamics of polymer translocation in the strong force regime 
by recasting the problem into solving a differential equation with a moving 
absorbing boundary.  For the total translocation time, $\tau_{\rm tr}$, our 
simple mean-field model predicts that $\tau_{\rm tr}\sim$ (number of 
monomers)$^{1.5}$, which is in agreement with the exponent found in 
previous simulation results. Our model also predicts intricate dependencies 
of $\tau_{\rm tr}$ on the variations of the pulling force and of the 
temperature.
\end{abstract}
\pacs{82.35.Lr,83.50.-v, 87.15.He, 05.40.-a} 

\maketitle


\section{Introduction}
\begin{figure}
\caption{A schematic of the translocation process. The bottom-most monomers 
are cleared first and the resulting deviation in concentration from the 
equilibrium condition induces a pressure gradient that 
drives the monomers at the interface, a distance $R(t)$ away from the pore, to the bottom.}
\label{main_pic}
\begin{center}
\includegraphics[scale=.4]{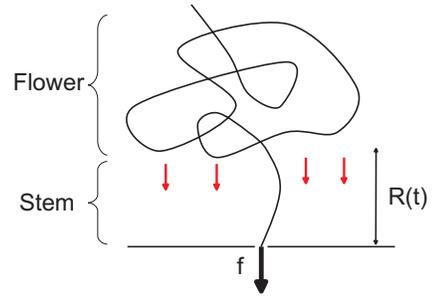}
\end{center}
\end{figure}

Understanding of polymer adsorption and translocation has important 
technological and biological significance. Besides well-known 
applications such as adhesion and coating, 
adsorption is responsible for facilitating breathing in the lungs 
\cite{Lung_surf} and translocation for the mechanism of transporting DNA 
and RNA across nuclear pores (e.g., see \cite{Newmeyer88}). 
Although usually viewed as two different phenomena, adsorption and 
translocation  are in fact very similar as 
both processes may be seen as having one end or both ends of a monomer 
chain pulled to an adsorbing surface. Indeed, simulation results have 
indicated that for a Rouse chain, both the total translocation time, 
$\tau_{\rm tr}$, and the total adsorption time $\tau_{\rm ad}$, scale like 
$M^{1.5}$ where $M$ is the initial number of monomers in the polymer 
globule \cite{Shaffer94, Grosberg06}.  Recently, it has also been 
recognised that many adsorption and translocation phenomena are in the 
strong force regime \cite{O'Shaughnessy05, Newmeyer88}, namely, $bf$ is at 
least a few times of $k_BT$ where $b$ is the bond length between connected 
monomers and $f$ is the effective force exerted on the monomers in 
adsorption or in translocation. This suggests that polymer adsorption or 
translocation is likely to be a nonequilibrium process. Coupled with the 
well-known observations of ageing  \cite{Frantz91} and glassy behaviour 
\cite{Chakraborty91, Chakraborty92, Srebnik96} in adsorbed polymer layer, 
the need for a better understanding of the dynamics of polymer adsorption 
and translocation is in order.
Here, we present a mean-field model that describes the dynamics of a 
polymer globule under translocation in the strong force regime. Letting the 
direction of the translocation process be in the negative $z$ direction, we denote the 
number density of monomers in each $xy$-plane along the $z$ axis by 
$\phi(t,z)$. The model is mean-field in the sense that the fluctuations in 
the $x$ and $y$ directions are averaged over. In the strong force regime, 
it is expected that the polymer will quickly adopt a ``stem-flower'' type 
configuration (c.f. Fig.~\ref{main_pic}) \cite{Brochard-Wyart95}.
We thus set up a differential equation with the bottom of the ``flower'' as 
a moving absorbing boundary. The differential equation based model can be 
numerical solved very efficiently. As our model retains all spatial 
information along the $z$-axis, it provides better characterization of the 
translocation process. For instance, it allows for the determination of the 
temporal evolution of the centre of mass for the portion of the polymer to 
be translocated.
Our model confirms the scaling law: $\tau_{\rm tr} \sim M^{1.5}$, found in 
simulation results \cite{Grosberg06}. We also obtain novel quantitative 
predictions concerning how $\tau_{\rm tr}$ would depend on the pulling 
force and thermal energy.

\section{Background}
We are interested in the strong force regime, i.e., the effective force 
involved is more than a few times of $k_BT/b$ where $b$ is the 
monomer-monomer bond length. In  the case of 
adsorption, this translates to having the adsorption energy, $\epsilon$, to 
be more than a few $k_BT$. 
In this regime, the polymer will be pulled strongly onto the surface and 
the time scale will generally be much quicker than the whole polymer 
relaxation time, $\tau_{\rm r} = \frac{m \gamma M^2 b^2}{3 \pi^2 k_BT}$ 
\cite{deGennes}, where $M$ is the number of monomers in the polymer, $m$ is 
the mass of each individual monomer, $\gamma$ is the monomer-solution 
collision frequency. The process is thus far from thermal equilibrium and 
local relaxation dominates. Our investigation is therefore fundamentally 
different from much of the earlier works
on polymer adsorption at equilibrium \cite{deGennes, 
deGennes81, Eisenriegler82}. 
Nonequilibrium polymer adsorption dynamics has also recently received much 
attention and most studies focused on the scaling for the adsorption time, 
$\tau_{\rm ad}$, which for a Rouse chain is found to scale like $M^{1.5}$ 
\cite{Shaffer94, Ponomarev00}. For further information on adsorption, we 
refer the readers to a recent review by  O'Shaughnessy and Vavylonis 
\cite{O'Shaughnessy05}. In the case of polymer translocation, interest in 
the physics community is comparatively more recent and most early  studies 
have focused on
the low force regime, in which the relaxation time is shorter than the 
process of translocation. This allows for the use of the Fokker-Planck 
equation description \cite{Sung96, Park98} or the nucleation theory 
\cite{Muthukumar99, Muthukumar01}. The validity of the above formalism has
been questioned in \cite{Chuang01} as it is argued that the relaxation time
and translocation time are of the same order of magnitude. The authors 
further demonstrate the existence of anomalous dynamics in translocation 
through simulations and scaling argument. Anomalous dynamics in forced 
translocation was also studied in \cite{Kantor04} and further 
explored in \cite{Dubbeldam07} with the use of fractional 
diffusion equation. More recently, the total translocation time for a Rouse 
chain, $\tau_{\rm tr}$, is investigated in \cite{Grosberg06} where
the authors argue that the pulling force would only affect a small portion 
(a ``fold'') of the polymer  at a time and starting from this assumption,   
$\tau_{\rm tr}$ is found by scaling argument to scale like $M^{1.5}$. 
However, Sakaue argued in \cite{Sakaue07} that the ``folds'' picture may
only be correct when $bf/k_BT > M^{1/2}$. In the paper,
the author treats the dynamics of translocation as a tension propagation 
problem and by assuming that each blob is at equilibrium locally, a 
differential equation governing the temporal evolution of $M$ under 
translocation is formulated and then solved numerically. The approach is
very similar in spirit to ours although there is one key difference:  we 
treat the thermal diffusion and applied force separately while Sakaue group 
them together in the form of an effective force: $\tilde{f} = fR_0/k_BT$,
where $R_0$ is the initial radius of the polymer globule. 
In terms of predictions, for a Rouse chain in the strong pulling regime, 
our model and Sakaue's model both give $\tau_{\rm tr} \sim M^{1.5}$, in 
agreement with simulation results \cite{Grosberg06}. On the other hand, our 
model indicates a much more complex relationship for $\tau_{\rm tr}$'s 
dependencies on $f$ and $k_BT$. In particular, we find that the scaling 
law: $\tau_{\rm tr} \sim f^{-1}$ is only true when $f/k_BT \rightarrow 
\infty$, and that $\tau_{\rm tr}$ is found to depend non-monotonically on 
the thermal energy.

\section{A discussion on scaling}
\label{scaling}
For the problem at hand, the dimensionful parameters are: $f, k_BT, b, 
\gamma$ and $m$,  with $M$ as the only dimensionless parameter. If we let 
the total translocation time, $\tau_{\rm tr}$, be given by the function: 
$\phi(f, k_BT, b, \gamma,  m,M)$. We can invoke the intuitive $\Pi$-theorem 
\cite{Pi} to transform the functional dependency into the following form:
\beq
\tau_{\rm tr} = \tau_0\Phi \left( 
\frac{bf}{k_BT},
\frac{k_BT}{mb^2\gamma^2}, b, \gamma, m, M
\right)
\eeq
where $\tau_0 = \frac{m\gamma b}{3\pi^2 k_BT}$ is the single monomer 
diffusion time \cite{deGennes}, and $\Phi$ is now dimensionless and as 
such, it can only depend on the first two and the last parameters, i.e.,
\beq
\label{Pi-th}
 \tau_{\rm ad} = \tau_0 \Phi \left(\frac{bf}{k_BT},
\frac{k_BT}{mb^2\gamma^2},M \right) \ ,
\eeq
The above equation is exact except for, of course, the fact that we do not 
know what $\Phi$ is. 

In \cite{Grosberg06}, the author's ansatz for the form of $\Phi$ is:
\beq
\Phi \left(\frac{bf}{k_BT},
\frac{k_BT}{mb^2\gamma^2}, N \right)
= {\rm const.} \times \frac{k_BT  }{bf}N^{3/2}\ .
\eeq
Namely, it is assumed that the second argument in $\Phi$ is redundant. 
There is no physical reasoning for this particular ansatz. Indeed, we find in 
this paper that all three arguments affect $\Phi$ independently even in the 
range $10 \leq bf/k_BT \leq 500$, which is not described in 
\cite{Grosberg06,Sakaue07}.

\begin{figure}
\caption{Temporal evolution of $\phi(t,z)$ with $m\gamma=1$, $b=1$, $f=10$, 
$D=1$ and $M=500$. The times are in units of $\tau_0$. Notice the gradual 
lengthening of the stem attached to the pore at $z=0$.}
\label{temporal}
\begin{center}
\includegraphics[scale=.5]{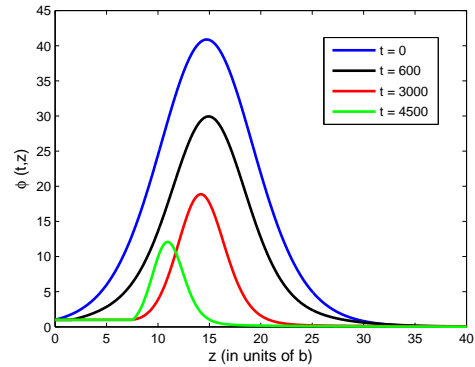}
\end{center}
\end{figure}
\begin{figure}
\caption{Some properties of $\phi(t,z)$ with parameters as defined in the 
caption of Fig.~\ref{temporal}. (a) Center of mass for the portion of 
polymer awaiting translocation (in units of $b$). (b) Number of monomers, 
$M(t)$. (c) The length of stem (in unit of $b$).}
\label{number}
\begin{center}
\includegraphics[scale=.5]{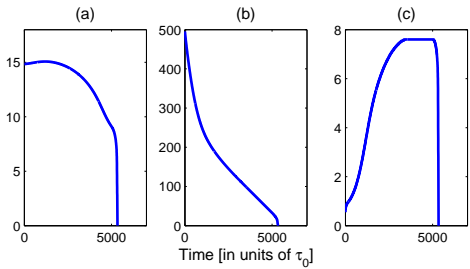}
\end{center}
\end{figure}

\section{A mean-field model}
Before we move on to describing our model, we remark that 
in this paper, we ignore all interactions from monomers that have passed 
through the pore. Namely, we assume that the pulling force is strong enough 
that the chain on the other side of the pore does not have any effect on 
the chain being translocated. The same assumption is made in 
\cite{Grosberg06,Sakaue07} and it renders our analysis more similar to the 
phenomenon of adsorption and so the findings here may be relevant to both 
phenomena.

The basic observation concerning translocation is that as the first monomer 
of a polymer globule is being dragged through the pore (c.f. 
Fig.~\ref{main_pic}),   the monomer number density at the bottom of the 
coil gradually decreases. This imbalance in monomer concentration from the 
equilibrium condition will induce an osmotic pressure that pushes the 
globule towards the surface. As a result, it accelerates the adorption 
process. 
Starting with this observation, we let $\phi(t,z)$ be the expected number 
of monomers at time $t$ and position $z$. In other words, $\phi(t,z)$ has 
dimension length$^{-1}$. Denoting the total number of monomers by $M$ and 
the center of mass by $\bar{z}$, we write $\hat{\phi}_{M,\bar{z}}$ as the 
monomers distribution at equilibrium. Given any other distribution, $\phi$, 
we assume that the osmotic pressure to be proportional to the difference 
between the current distribution and the distribution at equilibrium: $\phi 
- \hat{\phi}_{M,\bar{z}}$ (c.f. Ch.~VII in \cite{deGennes}).
In other words, if we ignore adsorption for the time being, the temporal 
equation on the distribution is: 
\beq
\frac{\pp \phi}{\pp t} = D \nabla^2  \left(\phi - 
\hat{\phi}_{M,\bar{z}}\right)
\eeq
where $D$ is the diffusion constant and is assumed to be  $k_BT/m\gamma$. 
We note that as a deterministic model, the above equation does not model 
diffusion of the whole molecules, i.e., $\bar{z}$ does not vary and as 
such, the model is meant to present the dynamical behaviour at short time 
in comparison to the whole globule relaxation time, $\tau_r = M^2\tau_0$. 
This assumption is consistent with the parameter set we study here as 
$\tau_{\rm tr}$ is always less than 10 percents of $\tau_{\rm r}$.

We now incorporate translocation into the model. If the monomers are not 
connected, the adsorption process may be modelled as a fixed absorbing 
boundary in the diffusion equation. But since the monomers are connected 
and as such the pulling force can propagate through the chain, the 
differential equation above becomes a moving boundary problem and we have 
the following model equation:
\beq
\label{diff_eq}
\frac{\pp \phi}{\pp t} = \left\{
\begin{array}{ll}
-\frac{f}{ m \gamma bR}   & , \ {\rm for} \ z = R(t)
\\
D\nabla^2 \left(\phi - \hat{\phi}_{M,\bar{z}} \right) & ,\ {\rm for} \  
z>R(t).
\end{array}
\right.
\eeq
where $R(t) = \max [z: \phi(t,z)< b \ {\rm and} \ 0\leq z \leq 
\bar{z}(t)]$, and $f$ is the pulling force. We also maintain that 
$\phi(z,t) =1$, for $0 \leq z <R(t)$, which represents the stem connecting 
the pore and the flower (c.f. Fig.~\ref{main_pic}).
In the above equation, $R(t)$ is the moving absorbing boundary with a 
constant absorbing rate $-\frac{f}{m \gamma bR}$. The rate equation can be 
obtained from the force-velocity equation:
\beq
\label{force_diss}
-\frac{bf}{m  \gamma R} = v = b^2\frac{\pp \phi}{\pp t} \ .
\eeq
By dimensional analysis, we know that $\hat{\phi}_{M,\bar{z}}(z) \equiv
\hat{\phi}_{M',\bar{z}}((M'/M)^\nu z)$ where $\nu$ is 3/5 for a chain in 
good solvent and it is 1/2 for a chain in $\theta$ solvent \cite{deGennes}. 
In other words, if we know $\hat{\phi}_{M(0),0}$, we can obtain all the 
other distributions $\hat{\phi}_{M,\bar{z}}$ by simple translation and 
dilation.
 
In summary, we have constructed a differential equation model that depends 
purely on a set of constant parameters: $f,m,b,\gamma, k_BT$, and a static 
distribution $\hat{\phi}_{M(0),0}$ that can be determined  once and for 
all.
Eq.~\ref{diff_eq} is the main result of this paper and it can be 
numerically solved efficiently (c.f. Appendices A and B for simulation 
details). We will now focus on the various predictions made by our model on 
the Rouse chain.

\begin{figure}
\caption{Adorption times with respect $M$. It is found that $\tau_{\rm ad} 
\sim M^{1.5}$. }
\label{times}
\begin{center}
\includegraphics[scale=.4]{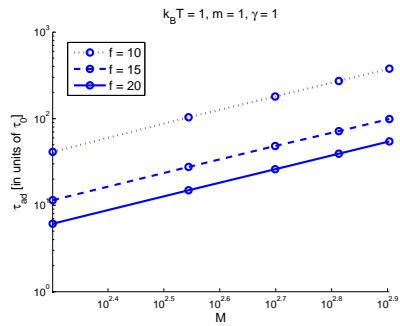}
\end{center}
\end{figure}
\section{Results and discussion}
Since the pulling force is strong, the diffusion process is not rapid 
enough to replenish the supply of monomer near the pore and so a stem forms 
naturally as shown in Fig.~\ref{temporal}. This renders an initial increase 
in separation between the center of mass and the pore as the bottom 
monomers are quickly translocated (c.f. Fig.~\ref{number} (a)).   
To understand the scaling law: $\tau_{\rm tr} \sim M^{1.5}$ (c.f. 
Fig.~\ref{times}), one can imagine the scenario where $k_BT \rightarrow 0$. 
In this situation, the globule is completely frozen throughout the 
translocation process. Now, since the initial size of the polymer globule 
is $\sim b\sqrt{M_0}$ where $M_0$ denotes the initial number of monomers, 
the stem will be of the same order of magnitude in length for most of the 
translocation process. Therefore, as a first approximation, one can set up 
the following differential equation as in Eq.~\ref{force_diss}:
\beq
-\frac{f}{m  \gamma b \sqrt{M_0}} =  \frac{\pp M(t)}{\pp t} \ ,
\eeq
and the scaling law will then follow immediately. In other words, the 
scaling is  purely due to the fact that the monomers being dragged to the 
pore are on average a distance of $\sim b\sqrt{M_0}$ away. This is a much 
simpler explanation of the scaling law than that presented in 
\cite{Grosberg06}  and it highlights that the ``fold'' picture may not be 
necessary in explaining the scaling behaviour seen in single chain 
translocation \footnote{The paper investigates branched polymer 
translocation  as well and this criticism does not extend to that 
consideration.}.

\begin{figure}
\caption{Adorption time vs. pulling force. 
The results indicate that the scaling law: $\tau_{\rm tr} \sim f^{-1}$ is 
only true asymptotically as $k_BT \rightarrow 0$, and deviation from it can 
be observed even for $k_BT = 0.04$ and for the range $10 \leq f \leq 500$.
}
\label{force}
\begin{center}
\includegraphics[scale=.4]{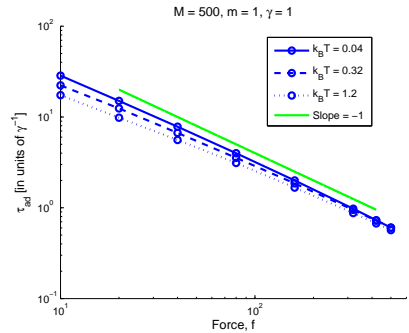}
\end{center}
\end{figure}

If the pulling force is increased, it is natural to expect that $\tau_{\rm 
tr}$ will decrease. This is indeed the case, but deviation from the 
expected scaling law: $\tau_{\rm tr} \sim f^{-1}$ can be seen even for the 
range $10 \leq bf/k_BT \leq 500$ (c.f. Fig.~\ref{force}). This is different 
from the expectation described in \cite{Grosberg06, Sakaue07}. In fact, our 
results suggests that the above scaling only holds at the limit $f/k_BT 
\rightarrow \infty$ and as such, highlight the important role of the 
thermal energy.

\begin{figure}
\caption{Adorption time vs. thermal energy. Notice that as $f$ increases, 
$\tau_{\rm tr}$ can become non-monotonic with respect to $k_BT$ as shown by 
the peak indicated by the black arrow.
}
\label{thermal}
\begin{center}
\includegraphics[scale=.4]{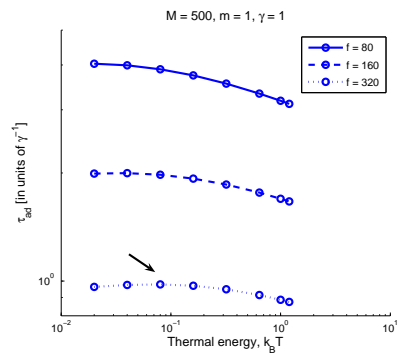}
\end{center}
\end{figure}

If the temperature is increased, the diffusion process (indicated by the 
red arrows in Fig.~\ref{main_pic}) induced by the osmotic pressure is 
facilitated and one would expect an decrease in $\tau_{\rm tr}$. 
Although this is generally the case, it is surprising to see the opposite 
trend at the low-temperature-high-force regime (c.f. Fig~\ref{thermal}). 
This counter intuitive feature may be understood by the fact that at low temperature, as the 
force is becomes large, the center of mass of the remaining 
polymer is driven away from the pore quickly (as shown in 
Fig.~\ref{number}) and this {\it escape} process is aided by
a slight increase in diffusion as the temperature is increased.

In conclusion, we have formulated a simple mean-field model for polymer 
translocation that captures the effect of local diffusion. Our model is 
capable of confirming the scaling law: $\tau_{\rm tr} \sim M^{1.5}$, as 
observed in simulations \cite{Grosberg06}, and predicts an intricate 
$\tau_{\rm tr}$'s dependencies on the pulling force and the thermal energy. 
Our work thus signals an interesting new territory that awaits exploration.

\begin{acknowledgements}
The author thanks the Glasstone Trust (Oxford) and Jesus College (Oxford) 
for financial support. 
\end{acknowledgements}

\appendix
\section{Fitting for $\hat{\phi}_{M,\bar{z}}$}
\begin{figure}
\caption{Fitting by the expression shown in Eq.~A1.}
\label{fitting}
\begin{center}
\includegraphics[scale=.4]{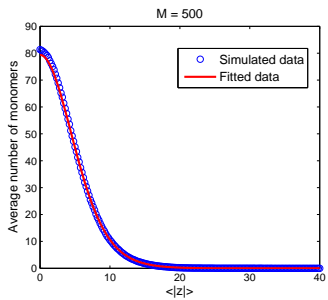}
\end{center}
\end{figure}

For a Rouse chain in three dimensions with $b=1$, we find that the 
following function is a good approximation for
$\hat{\phi}$:
\beq
\hat{\phi}_{M,\bar{z}}(z) = \frac{\sqrt{M}}{2 }\exp \left[ P \left( 
\frac{z-\bar{z}}{\sqrt{M}} \right) \right]
\eeq
where $P(x)$ is 
\beq
 -0.6x^{10} + 4x^8 - 9.48x^6+11.21x^4-11.94x^2+1.27 
\eeq
for $|x|<1.44$ and $\hat{\phi}_{M,0}(x)=0$ otherwise (c.f. 
Fig.~\ref{fitting}). This functional form is used in our numerical 
integration although this approximation step by an closed form function is 
not necessary. Instead, one can formulate a lookup table for $\nabla^2 
\hat{\phi}_{M,\bar{z}}$ from sampling. 

\section{Details on simulation method}
In solving the differential equation Eq.~\ref{diff_eq}, we employ the 
finite element method. Namely, we denote $\phi(t_n,z_j)$ by $\phi^n_j$ 
where $t_n$ and $z_j$ are the grid points on time and on position. In our 
simulations, $\triangle t = 0.001\gamma^{-1}$ and $\triangle z = 0.05b$.
Specifically, our algorithm is as follows:
\begin{enumerate}
\item
Given $M$ and a time grid and a position grid with spacing $\triangle t$ 
and $\triangle z$, set $n=0$, $\bar{z}=0$, and for all $j$, set $\phi^0_j = 
\hat{\phi}_{M,0}(z_j)$ where $\hat{\phi}_{M,0}$ is given in Appendix A.
Let $p_0=\max[j : \phi^0_j<1 \ {\rm and} \ z_j < \bar{z}]$, set $s 
=z_{p_0}$, $R = \triangle z$ and $v= \frac{bf}{mR \gamma}$.
\item
For $j>p$, set $\phi^{n+1}_j$ as
\[
 \phi^n_j + D {\triangle t}  
 \left( \frac{\phi^n_{j+1}-2\phi^n_j+\phi^n_{j-1}}{\triangle z^2} - 
\nabla^2 \hat{\phi}_{M,\bar{z}} \right) - \triangle t v\delta_{j,p+1}\ .
\]
\item
For $p_0 \leq j \leq p$, set $\phi^{n+1}_j$ to 1.
\item
Set $M$ as $M-v\triangle t$ and $\bar{z} = \sum_j z_j \phi_j^{n+1}/\sum_j 
\phi_j^{n+1}$.
Renormalise $\sum_j \phi^{n+1}_j$ to $M$ by re-scaling $\phi^{n+1}_j$.
\item
Let $p=\max[j : \phi^{n+1}_j<1 \ {\rm and} \ z_j < \bar{z}]$, set $R = 
z_p-s+ \triangle z$ and $v= \frac{bf}{mR \gamma}$.
\item
If $M<1$, stop; otherwise, increment $n$ by 1 and go back to 2. 
\end{enumerate}

\bibliography{new_adsorption}

\end{document}